\documentclass[a4paper,preprint,endfloats]{revtex4}
\usepackage{graphicx}
\usepackage{courier}
\usepackage{amssymb,amsmath}
\makeatletter
\def\@dotsep{4.5}
\makeatother
\begin{document}

\title{COLLISIONAL QUENCHING AT ULTRALOW ENERGIES: CONTROLLING EFFICIENCY WITH INTERNAL STATE SELECTION}
\author{S. Bovino, E. Bodo and F.A. Gianturco\footnote{Corresponding author; e.mail
address: fa.gianturco@caspur.it; fax: +39.06.49913305.}}
\affiliation{Department of Chemistry and CNISM, University of Rome
La Sapienza, Piazzale A. Moro 5, 00185 Rome, Italy}

\begin{abstract}
Calculations have been carried out for the vibrational quenching
of excited H$_2$ molecules which collide with Li$^+$ ions
at ultralow energies. The dynamics has been treated exactly using
the well known quantum coupled-channel expansions over different initial
vibrational levels. The overall interaction potential has been obtained
from the calculations carried out earlier in our group
using highly correlated ab initio methods. The results indicate that specific
features of the scattering observables, e.g. the appearance of
Ramsauer-Townsend minima in elastic channel cross sections and the
marked increase of the cooling rates from specific initial states,
can be linked to potential properties at vanishing energies (
sign and size of scattering lengths) and to the presence of either
virtual states or bound states. The suggestion is made that by
selecting the initial state preparation of the molecular partners,
the ionic interactions would be amenable to controlling quenching
efficiency at ultralow energies.
\end{abstract}

\maketitle

\section{Introduction}
Manipulation of atomic and molecular systems at ultralow energies
has completely changed the research landscape of experimental and
theoretical molecular phenomena. Thus, intense efforts have been
directed to the study of new properties of Bose-Einstein condensates
\cite{1,2} and to analyze the quantum degeneracy features of Fermi
gases like $^{40}$K \cite{3} and $^{6}$Li \cite{4}, since ultracold
quantum matter containing fermions has exhibited a new range of
unexpected properties. The occurrence of Bose-Einstein condensation
and of Fermi degeneracy in diluted gases typically requires
temperatures between 1nK and 1$\mu$K, although new quantum effects
can begin to show up already around a few mK, where the Broglie
wavelengths often become large compared to the available atomic and
molecular dimensions. Under such special environmental conditions
the relevant collisions between partners, be them atomic or/and
molecular, become fully quantal and are chiefly guided by the
long-range interactions: the regions just below 1mK are usually
opening up the ultracold collisional regimes.

Over the last few years, with the increase of the interest on
molecular systems in the ultracold regime, and on the processes 
taking place in cold traps or in Coulomb crystals involving both neutral and ionic molecules, there
has been a shift of focussing onto the details and features of
possible means one can employ to control ultracold molecular quantum systems. The latter, in fact, show a much
richer structure of internal energy levels with respect to atoms and offer more
possibilities for quantum control. For example dipolar gases are predicted to
exhibit new features which have likely applications in quantum
computing \cite{5,6} and the possibility of manipulating their
reactive and cooling efficiency from selected initial states still
remains an intriguing process which is clearly linked to their
structural features.

Furthermore, the direct process of cooling molecules from initial
room temperature conditions by using precooled atomic gases is one
of the mayor ways in which such special, ultracold  systems could be
prepared \cite{7}. It therefore becomes important to be able to know
just how efficient the cooling processes can be and how much they
are influenced by the initial preparation of the molecular targets.
The computational modelling of such assessments is, however, a
rather complicated affair where one needs to know reliably the
overall range of interaction between molecular partners, or between
molecules and cold atoms, the detailed dynamical coupling matrix
elements between that interaction and the rotovibrational manifold
of at least one molecular partners and, finally, one should be able
to solve exactly the coupled-channel (CC) quantum dynamics at the
cold and ultracold temperature regimes. The last requirement,
however, become easier to achieve under those special conditions
since the quantum suppression effects strongly reduce the number of
contributing angular momenta during the scattering event \cite{8,9}.
Furthermore, the marked disappearance of quantum interference
between trajectories (as only the s-wave contribution dominates the
scattering S-matrix) enhances the nanoscopic features of each
collision and suggests that ultimate control of the scattering
attributes could be more easily detected when dealing with cold and ultracold
scattering experiments \cite{10} than in the case of room temperature dynamics.

The aim of the present work is therefore that of showing, in a
specific molecular system where reactive channels can also become
open at higher energies, the effect on scattering observables and
on inelastic quenching cross sections of starting the collisional
process with the molecular partner (the H$_2$ molecule) prepared in
a specific vibrational level. 
We had already studied the neutral Li$_{2}$+He molecular system in the case of 
non reactive events at ultralow energies \cite{11}
and had found a marked dependence of cross section magnitudes on the
initial vibrational channel, although the reaction examined involved
a rather weak interaction among neutral partners. In the present
study we shall analyse instead a collision process in an ionic system
\begin{equation}
\rm
Li^++H_2(\emph{v})\leftrightarrows Li^++H_2(\emph{v}'<\emph{v})
\end{equation}
to see the dynamical behaviour when  a stronger potential energy
surface (PES) is involved, and a PES for which a reactive channels also opens up
for $v>11$
\begin{equation}
\rm
Li^++H_2(\emph{v}>11)\leftrightarrows LiH^++H
\end{equation}
although we shall not be considering in the present work
the additional contributions from such channels. We have already analyzed
in a preliminary report \cite{14} the quenching cross sections of reaction (1) 
for the lowest three vibrational states of H$_2$. The aim of the present work is 
to extend the study to the full range of accessible initial states, thereby providing 
a more complete picture on the controlling capabilities of initial state preparation 
in molecular encounters.

The following Section 2 will therefore outline our scattering
equations and their behaviour at ultralow energies, together with
sketching the general features of the involved potential energy
surface. The next Section 3 will present our results while
Section 4 will summarize our conclusions.

\section{Interaction Forces and Cold Dynamics }
The present calculations employ the adiabatic electronic ground
state of the LiH$_2^+$ complex at collision energies well below its
barrier to the reaction of eq. (2)
\noindent so that we can assume that the inelastic channels of the
process (1) are the only ones which are open at the considered
energies: the reaction of eq. (2) is, in fact, endothermic by about 4.4 eV
due to the lower binding energy of LiH$^+$ with respect to H$_2$. It
therefore shows a purely uphill route in going to the LiH$^+$
formation, with an absolute minimum energy geometry of the Li$^+$H$_2$ complex in C$_{2v}$
orientation. The details of the full reactive PES, together with
those on the quality of the employed ab initio method to yield the
raw points of the surface, have been discussed before \cite{12,13}
and will not be repeated here. Suffice it to say that on the
reagents' region of the Li$^+$ ion approaching the H$_2$ molecule,
the corresponding minimum energy path exhibits a well of about 2300
cm$^{-1}$, which reduces to $\sim$ 650 cm$^{-1}$ for the collinear
orientations. That part of the interaction is dominated by
the charge interacting with the molecular permanent quadrupole and
by the spherical and non-spherical polarizabilities of the H$_2(v)$
molecule. To briefly remind readers of its general features, the collinear 
profile of the reactive path is sketched in figure 1,
where we also report the relative locations of the first eleven
vibrational states of H$_2$ existing below the barrier to LiH$^+$
formation, a process which we therefore will not be considering to
occur at the ultralow collision energies examined in the present
study.

\begin{figure}
\begin{center}
\includegraphics[width=0.9\textwidth]{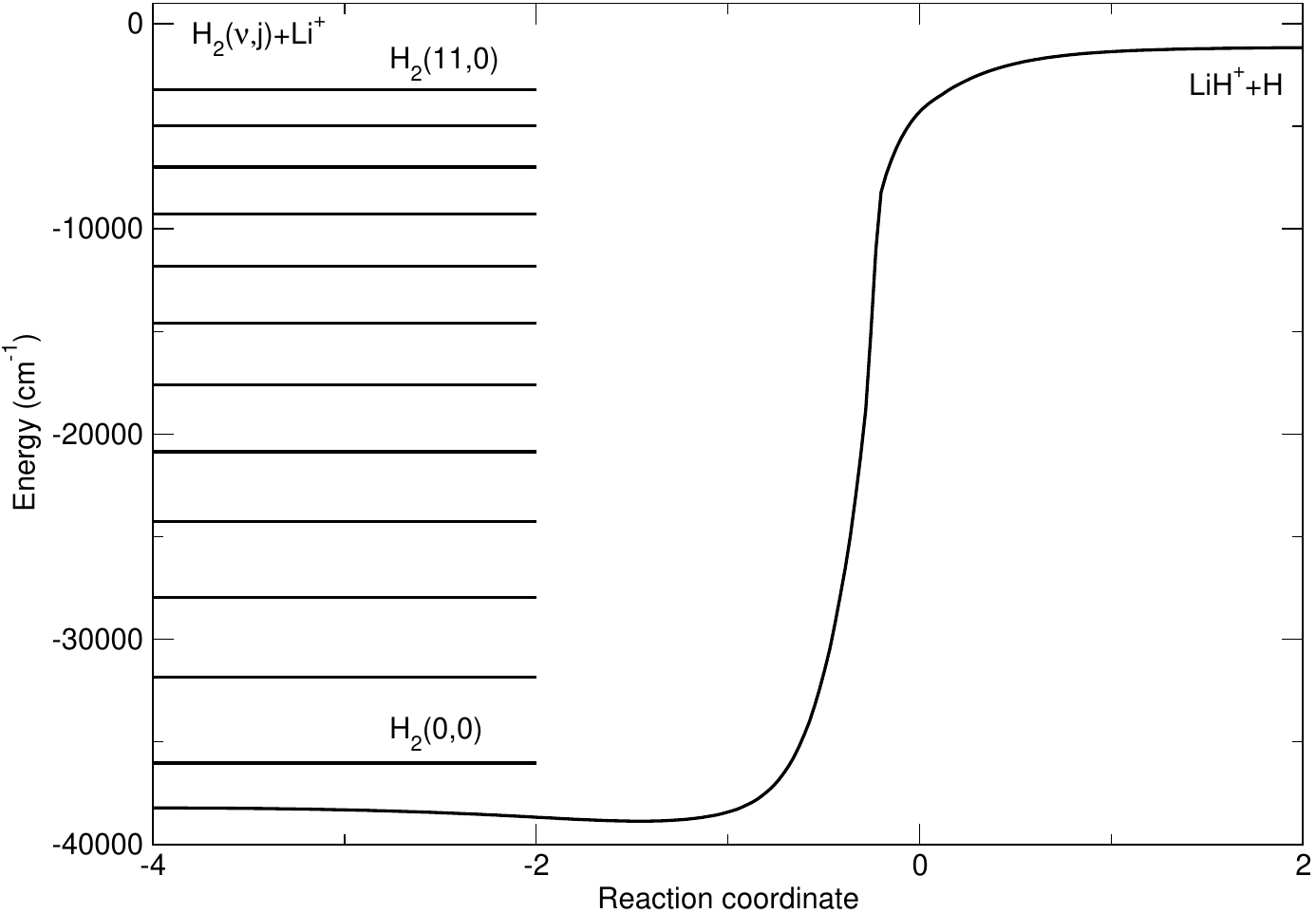}
\caption{Collinear minimum energy path profile of the Li$^++\mathrm{H_2}$. The molecular
energy levels accessible below the reactive barrier are also
reported (all energies in cm$^{-1}$ and distances in
{\AA}). Adapted from Ref.  \cite{14}.}
\label{figure1}
\end{center}
\end{figure}

One clearly sees in the figure that all collisions starting with
vibrationally excited hydrogen molecules up to the 12th vibrational
level, and occurring at vanishing energies, will be below
barrier and will thus only access the inelastic channels of eq. (1). 
One should also keep in mind that the processes,
which we shall be analyzing, occurring very closely to the near-zero
energy region of the upper rim of the interaction potentials, will
depend rather crucially on the efficiency of the potential couplings
between target levels and therefore on the shape of the matrix
elements given by convolutions over target asymptotic vibrational
states $\varphi _i (r)$
\begin{equation}
V_{ij}(R,\vartheta)=\langle \varphi _j|V(R,r,\vartheta )|\varphi
_i\rangle
\end{equation}
which will therefore appear in the corresponding familiar CC formulation 
of the collision problem discussed below.

The vibrational wave functions and the vibrational levels of H$_2$
$\varphi _i=\varphi _{vj}$ are obtained by solving the
Schr\"{o}dinger equation for the diatom using the potential of ref. \cite{19}. 
The corresponding coupled-channel equations in
the space-fixed (SF) reference frame are then given by
\begin{equation}
\left\{\frac{d^2}{dR^2}+\mathbf{k}^2-\mathbf{V}-\frac{\ell
^2}{R^2}\right\}\mathbf{F}^J=0
\end{equation}

\noindent where, as usual, $[\mathbf{k}^2]_{ij}$ is the diagonal
matrix for the asymptotic (squared) wave vectors, the coupling
matrix $\mathbf{V}=2\mu \mathbf{U}$ contains the full potential
terms and $\ell ^2$ is the matrix representation of the square of
the relative orbital angular momentum: $[\ell ^2]_{ij}=\delta
_{ij}\ell _i(\ell _i+1)$. The matrix $\mathbf{F}^J$ holds the radial
solutions for each choice of the total angular momentum
$\mathbf{J}$, the dynamical constant of the motion for the present
system.

In the asymptotic region the solution matrix can be written as
\begin{equation}
\Psi (R)=\mathbf{J}(R)-\mathbf{N}(R){\cdot}\mathbf{K}
\end{equation}

\noindent where $\mathbf{J}(R)$ and $\mathbf{N}(R)$ are
diagonal matrices containing Riccati-Bessel and Riccati-Newmann
functions and from which we obtain $\mathbf{K}$ and then $\mathbf{S}$. The corresponding state-to-state superelastic cross
sections will be given by
\begin{equation}
\sigma_{vj,v'j'}(E_i)=\frac{\pi}{(2j+1)k_{vj}^2}\sum_{J}(2J+1)\sum_{l,l'}
|\delta _{lvj,l'v'j'}-S_{vj,v'j'}^{ll'}|^2
\end{equation}

The corresponding expansion of the elastic matrix element for $\ell$=0 in powers
of \emph{k} allows one to write \cite{8a,15}
\begin{equation}
S_{vj,v'j'}\simeq
1+2i\delta_{vj}(k)=1-2ik(\alpha_{vj}-\beta_{vj})=1-2ika_{vj}
\end{equation}

\noindent which makes us obtain the real $\alpha_{vj}$  and
imaginary $\beta_{vj}$ parts of the scattering length $a_{vj}$.
The corresponding elastic and inelastic parts of the total
scattering cross sections, in the same limit of $k \rightarrow 0$,
are given by
\begin{equation}
\sigma_{vj}^{el}=4\pi
|a_{vj}|^{2},\,\,\,\,\,\sigma_{vj}^{in.}=\frac{4\pi\beta_{vj}}{k}
\end{equation}

We can further derive the behavior of quenching rates for a selected
initial state, by relating it to the imaginary part of the
scattering length or, equivalently, to the inelastic quenching cross
section \cite{8a,15}
\begin{equation}
R_{vj}=\frac{4\pi\hbar}{\mu}\beta_{vj}
\end{equation}

\noindent At ultralow collision energies the features of the
scattering attributes as obtained from the computed S-matrix (i.e.
the size and sign of the complex scattering length for each initial
state) can tell us of the
existence for the corresponding scattering potential energy surface
of either zero-energy resonant states (virtual states) or of bound
states supported very near the threshold energies
depending on the sign of the real part of the complex scattering
length, $\alpha_{vj}$. The (complex) energy of these states is given by:
\begin{equation}
E=-\frac{\hbar^{2}}{2\mu|a_{vj}|^{2}}(\cos2\gamma_{{v}j}+i\sin2\gamma_{{v}j})=E_{{v}j}-\frac{i}{2}\Gamma_{{v}j}
\end{equation}

\noindent where $\gamma_{v j} = tan^{-1}\frac{\beta_{v j}}{\alpha_{v j}}$. 
Whenever a positive scattering length exists, one
can see it to increase to infinity as $E$ approaches zero
energy from below and to reappear at increasingly larger negative
values as $E$ moves on the second Riemann's sheet as a virtual
state of the ionic complex. The above quantities obviously depend on the
interaction potential and hence on the features of the coupling
matrix elements between vibrational levels involved in the quenching
process. This is particularly so with ionic interactions where the
long-range coupling in all channels could produce, even at ultralow
energies, contributions to the opacity functions beyond the more
usual s-wave contribution \cite{14,15,16}. It is one of the aims of
the present work to show that this is indeed the case in the present
system and that the presence near the potential top of either bound
states or virtual states strongly affects the quenching efficiency
of the ultracold H$_2$ molecules.

To give us a pictorial idea of the features of the coupling
potential terms between vibrational levels of the target molecule,
we report in figure 2 their radial shapes at two different
orientations and between  some of the vibrational levels: only
diagonal terms are shown in the two panels of that figure.

One notices from the figure that the overall potential strength increases as the molecule is vibrationally more excited: the repulsive walls at $0^{\circ}$  move outwards as $v$ increases due to the stretching of the molecular bond. The physical consequances of such feature on the behavior of the quenching cross section will be further discussed in the next section.
\begin{figure}
\begin{center}
\includegraphics[width=0.9\textwidth]{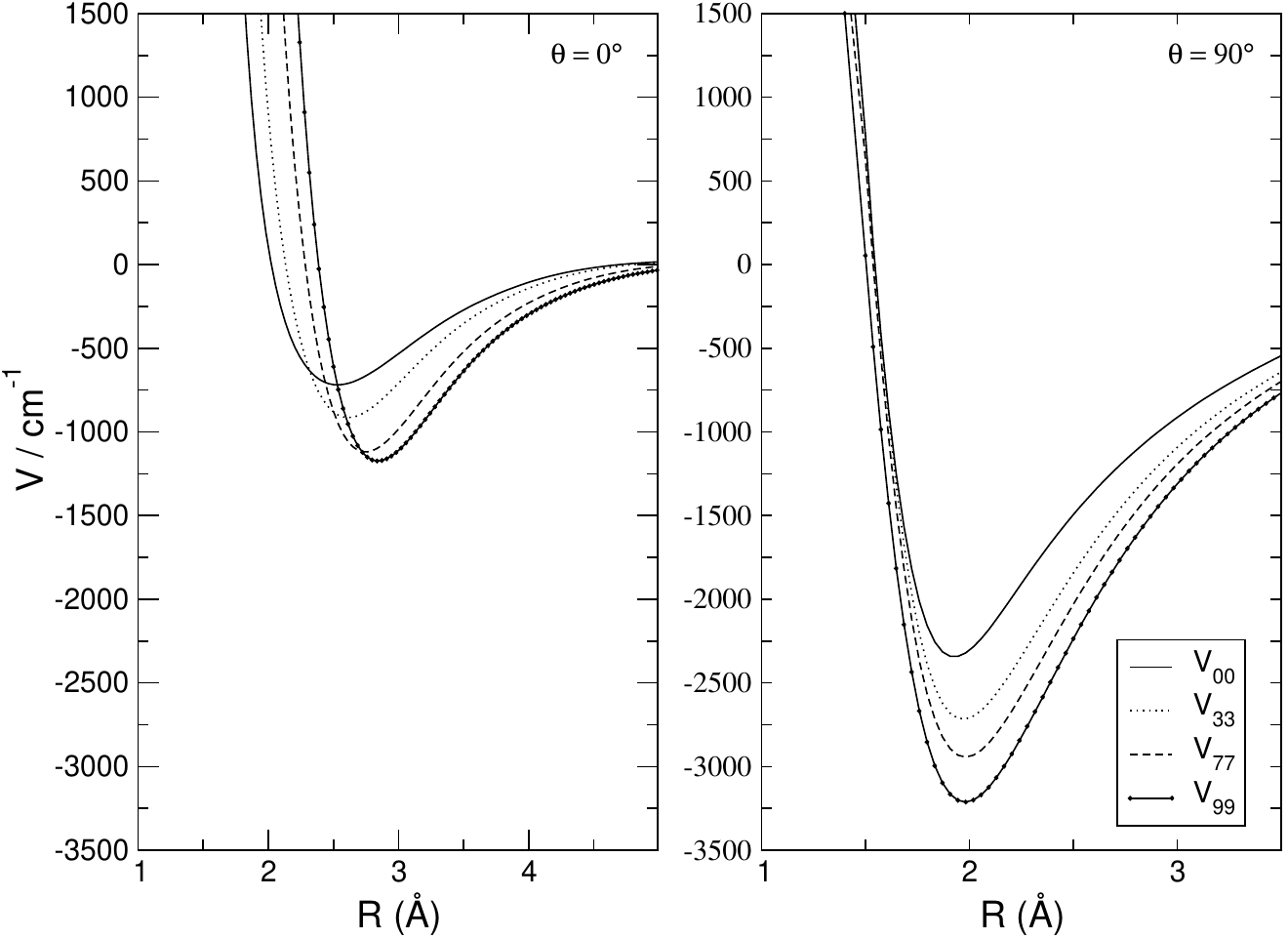}
\caption{Computed coupling matrix elements (diagonal terms) between
H$_2$ vibrational levels and for two different approaches of the
lithium ion. Left panel: collinear approach; right panel: T-shaped
approach.}\label{figure2}
\end{center}
\end{figure}

\section{Elastic and Quenching Collisions}
As mentioned before, the aim of the present study is to extend our 
earlier dynamical study \cite{14} on the
feasibility of controlling the quenching efficiency of vibrationally
excited H$_2$ molecules, when colliding at ultralow energies with a
simple molecular ion, to the full range of accessible vibrational levels in order to see which 
cross sections and rates exhibit some special behavior,
that leads to larger relaxation rates or to special features in the
elastic cross sections.

In order to describe the molecular target we have employed all the
open vibrational channels for a specific initial state and three closed
vibrational channels above them. Furthermore, for each vibrational
level, up to $j_{max}=12$ rotational channels were included in the
expansion for a maximum of about 70 asymptotic coupled channels.
The corresponding coupled equations were propagated from
the inner region out to 500 {\AA} and the values of total
angular momenta  needed for convergence ranged from $J$=0 (at the lowest energies) up to
$J$=9 (at the highest ones). 

The computed elastic cross sections as a function of collision
energy, on a log-log scale, are reported by figure 3 for each of the
initial vibrational states in which the H$_2$ molecule is being
prepared (the rotational initial state is $j=0$).

\begin{figure}
\begin{center}
\includegraphics[width=0.9\textwidth]{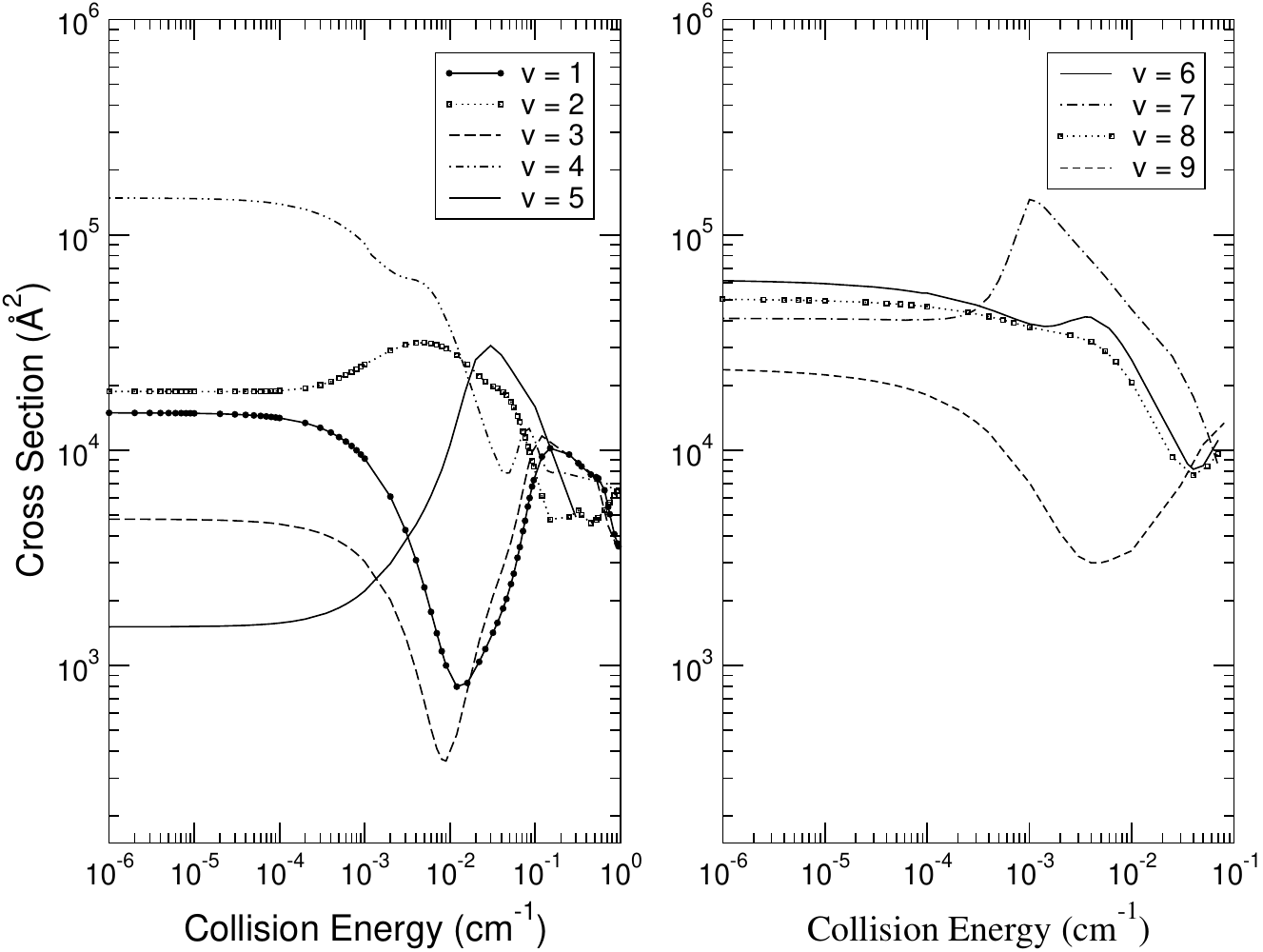}
\caption{Computed elastic cross sections for the Li$^++H_2(v)$
system as a function of collision energy and for a sequence of
vibrationally excited initial state of the molecule. See text for
computational details.}\label{figure3}
\end{center}
\end{figure}

From the data reported in that figure the following considerations
could be made:
\begin{itemize}
  \item at collision energies below \rm 10$^{-3}\mathrm{cm}^{-1}$ one notices the
  linear dependence of the elastic cross sections on the energy
  and their being constant in value below 10$^{-4}\mathrm{cm}^{-1}$: this
  means that we have reached the low-energy Wigner's law regime
  \cite{17}, whereby the $\ell$=0 contributions to the cross sections dominate the scattering and become
  energy-independent in the elastic channels.
  \item at the collision energies between 10$^{-2}$ and
  $2 \cdot 10^{-3}\mathrm{cm}^{-1}$ specific cross sections for the molecular
  target initially prepared in the $v$=1,3 and 9 vibrational states
  show a very marked drop. Such features are indicative of the presence of
  Ramsauer-Townsend (RT) minima, whereby the scattering process is
  mainly controlled by \emph{s}-wave and the corresponding eigenphase for the $\ell$=0
  channel vanishes \cite{18}. The \emph{s}-wave scattering length is
  negative and a virtual state can be present in the triatomic complex.
\end{itemize}

The fact that three different target preparations can lead to such
marked features of the elastic channels at energies around the mK
regime suggests that they could possibly be amenable to experimental
detection, thereby providing a rather sensitive test on the quality
of the ion-molecule interaction potential employed by our
calculations.

The differences which exist once the inelastic quenching cross
sections are examined, could be gleaned by looking at the
corresponding dominant partial wave contributions to inelastic
channels. The results reported in the four panels of figure 4 show,
in fact, the relative importance of  two  partial waves (J=0 and J=1)
for possibly selecting quenching cross sections for which  the H$_2$
molecule is prepared in a specific initial vibrational state ($v$,j=0)
and decays into the ($v$-1, j=2) states.

\begin{figure}
\begin{center}
\includegraphics[width=0.9\textwidth]{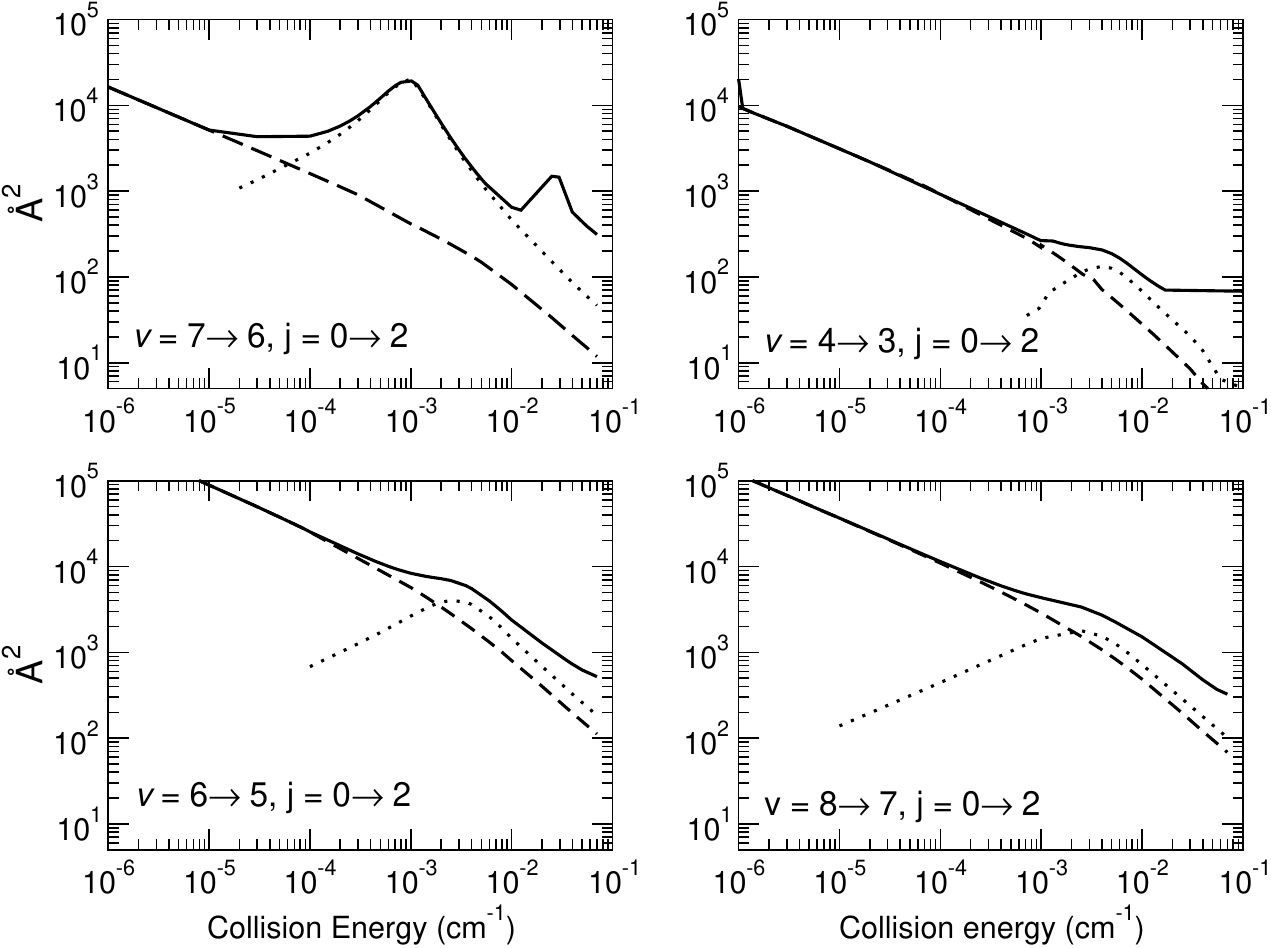}
\caption{Computed quenching cross sections for collisional
de-excitations from different initial states and for the ($\Delta v$=-1, $\Delta j$=+2) process. The dominant contributions of J=0
(dashes) and J=1 (dots) are shown for each integral cross section
(solid lines).}\label{figure4}
\end{center}
\end{figure}
One clearly sees that, for all the de-excitation processes presented
in the panels, the J=0 contributions are by far the
dominant ones as one reaches the Wigner regime of the cross
sections. On the other hand, in the collision energy range of the
10$^{-3}$ wavenumbers one clearly sees the very strong additional contribution from
the J=1 (\emph{p}-wave) channels that suggest the presence of additional
open channel resonances (dynamical trapping) induced by the strong
ion-molecule long-range contributions to the scattering process.

We have also found similar \emph{p}-wave contributions appearing for
ultra-cold collisions for H$_2$ molecules prepared in the $v$=1 and $v$=2
initial states only, and  discussed it in our previous preliminary report
\cite{14}: we show here that such resonant features indeed extend up the
vibrational ladder to the higher initial states presented by figure
4. With the same token, the features of the elastic channels
presented by figure 3 extend to the $v$=9 initial state of the
molecular partner the RT minimum appearance already discussed for
the $v$=1 and $v$=3 states \cite{14}.

The presence of marked \emph{p}-wave contributions to the quenching cross
sections around the 10$^{-3}$ cm$^{-1}$ energy regime will obviously
affect the efficiency of the quenching process and therefore will be
reflected in the corresponding behavior of the inelastic cross
sections.

The calculations reported by figure 5, in fact, show the  behavior of
the total inelastic cross sections (summed over all final ro-vibrational states) originating from a set of
selected initial vibrational states of the molecular partner (the rotational initial state is $j=0$): the
energy region goes from 0.1 wavenumber down to 10$^{-6}$cm$^{-1}$.

\begin{figure}
\begin{center}
\includegraphics[width=0.9\textwidth]{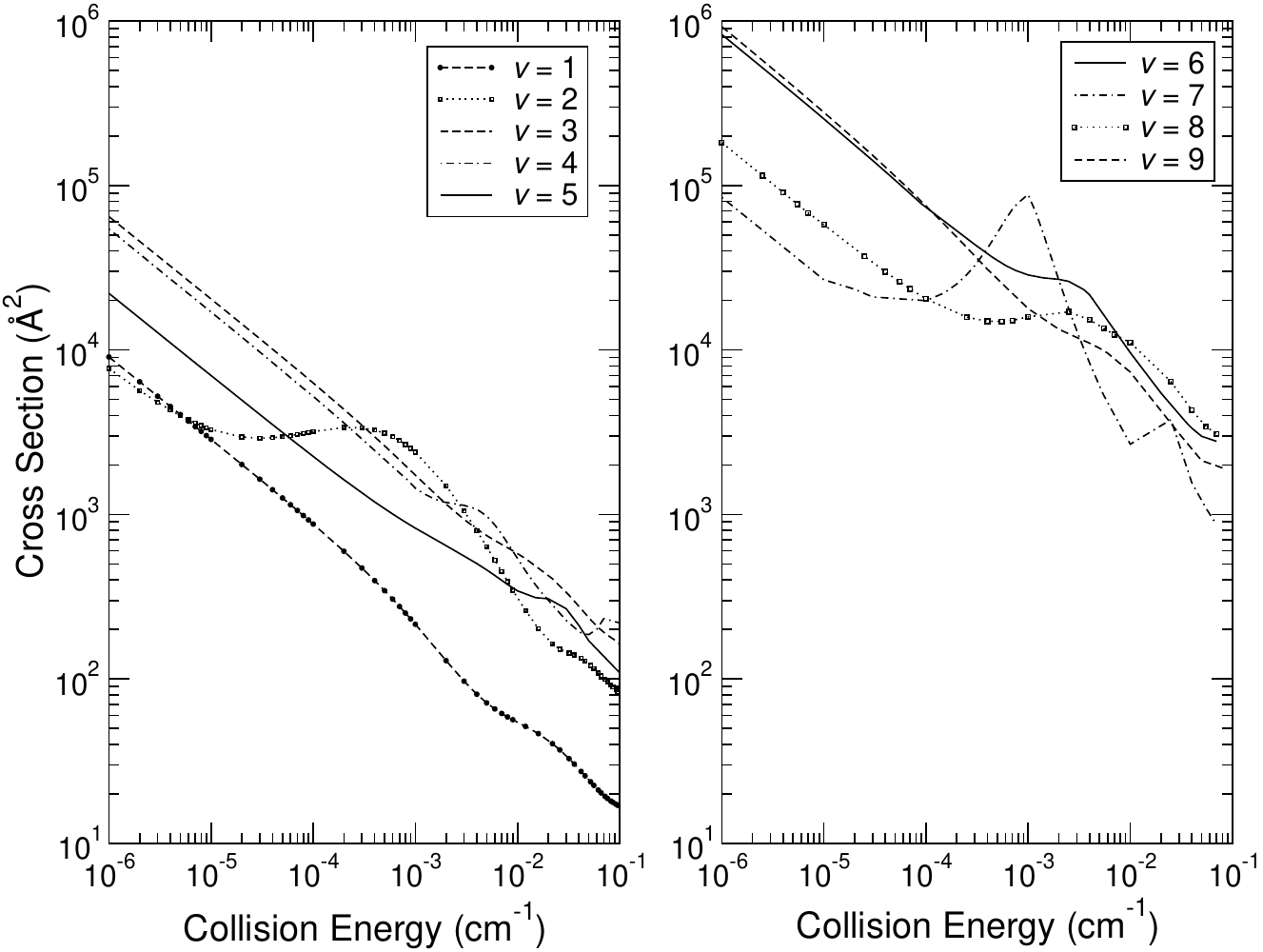}
\caption{Computed total quenching cross sections (vibrational) from
different initial ($v$,0) states of the H$_2$ molecule. See text for
details.}\label{figure5}
\end{center}
\end{figure}

The following comments could be made from a perusal of the reported
data:
\begin{itemize}
  \item the initial molecular preparation in the $v$=7, which had
  shown a strong \emph{p}-wave resonant contribution in figure 4, shows now the
  largest cooling efficiency in the milli-cm$^{-1}$ energy region;
  \item the initial levels 6, 8 and 4 show decreasing effects from
  \emph{p}-wave contributions (see figure 4). Hence, their quenching cross
  sections within the same range of energies also show decreasing
  efficiency in going from $v$=6 to $v$=8 and to $v$ =4;
\end{itemize}

The overall behavior of the scattering observables as a function of
the prepared initial vibrational state is given by the numbers
reported in Table 1. The quenching rates are given both as
asymptotic values at vanishing temperature and as convoluted values
around the millikelvin region (last column in Table). We clearly see
that those initial states where resonant features were detected from
\emph{p}-wave contributions (see figure 4) are also showing the largest
quenching rates across the same range of energies. This confirms
that ionic interactions can be capable, even at ultralow
temperatures, to cause dynamical trapping because of the $\ell$=1 barriers which
appear to give substantial contributions to the cross section at those energies. 
However, although the importance of such effects is a strong function of the selected 
molecular target, the present results indeed suggest that internal energy preparation at ultracold collisions
could be a possible path to controlling quenching efficiency of
ionic systems.
\begin{table}
\begin{center}
\begin{tabular}{c c c c c c c}
\hline \hline ~~$v$~~ & ~~$\alpha$(\AA)~~ & ~~$\beta$(\AA)~~ 
& E (10$^{-3}$ cm$^{-1}$) &
~~$\Gamma$ (ns)~~ & $R_{v}(T \rightarrow 0)$ (cm$^3\cdot$s$^{-1}$) &
$R_{v}$ (cm$^3\cdot$s$^{-1}$)\\
\hline \hline

0 & \ 36.90 & 0.0 & -8 & $\infty$ & 0.0 & -\\
1 & -34.60 & 0.22 & -8 & 23 & \ 1.12$\cdot10^{-11}$ & -\\
2 & \ 38.60 & 0.18 & -7 & 39 & \ 9.17$\cdot10^{-12}$ & -\\
3 & -19.46 & 1.65 & -28 & 0.5 & \ 8.42$\cdot10^{-11}$ & -\\
4 & \ 101.3 & 1.31 & -1 & 97 & \ 6.69$\cdot10^{-11}$ & -\\
5 & \ 8.30 & 0.81 & -152 & 0.09 & \ 4.14$\cdot10^{-11}$ & -\\
6 & \ 66.43 & 23.3 & -1 & 2 & 1.19$\cdot10^{-9}$ & 1.50$\cdot10^{-9}(10^{-3}K)$\\
7 & \ 53.65 & 1.99 & -3 & 9.5 & \ 1.01$\cdot10^{-10}$ & 1.70$\cdot10^{-9}(10^{-3}K)$\\
8 & \ 60.87  & 12.6 & -2 & 2.4 & \ 6.43$\cdot10^{-10}$ & 1.22$\cdot10^{-9}(10^{-2}K)$\\
9 &  -39.07  & 12.0 & -5 & 0.7 & \ 6.16$\cdot10^{-10}$ & -\\
\hline \hline
\end{tabular}
\end{center}
\caption{Real ($\alpha$) and imaginary ($\beta$) components of the
scattering lengths, energy locations (E) and lifetimes ($\Gamma$) of
the virtual and bound states, and quenching rates ($R_v$) from the
considered $v$ initial levels of H$_{2}$. The rates c(the rotational initial state is $j=0$)omputed in
the millikelvin regimes are shown in the last column for a few of
the initial target states.}
\end{table}

A summary of the  features which have been shown by the present
calculations is pictorially given by the results we collect in the
four panels of figure 6.

\begin{figure}
\begin{center}
\includegraphics[angle=-90,width=0.9\textwidth]{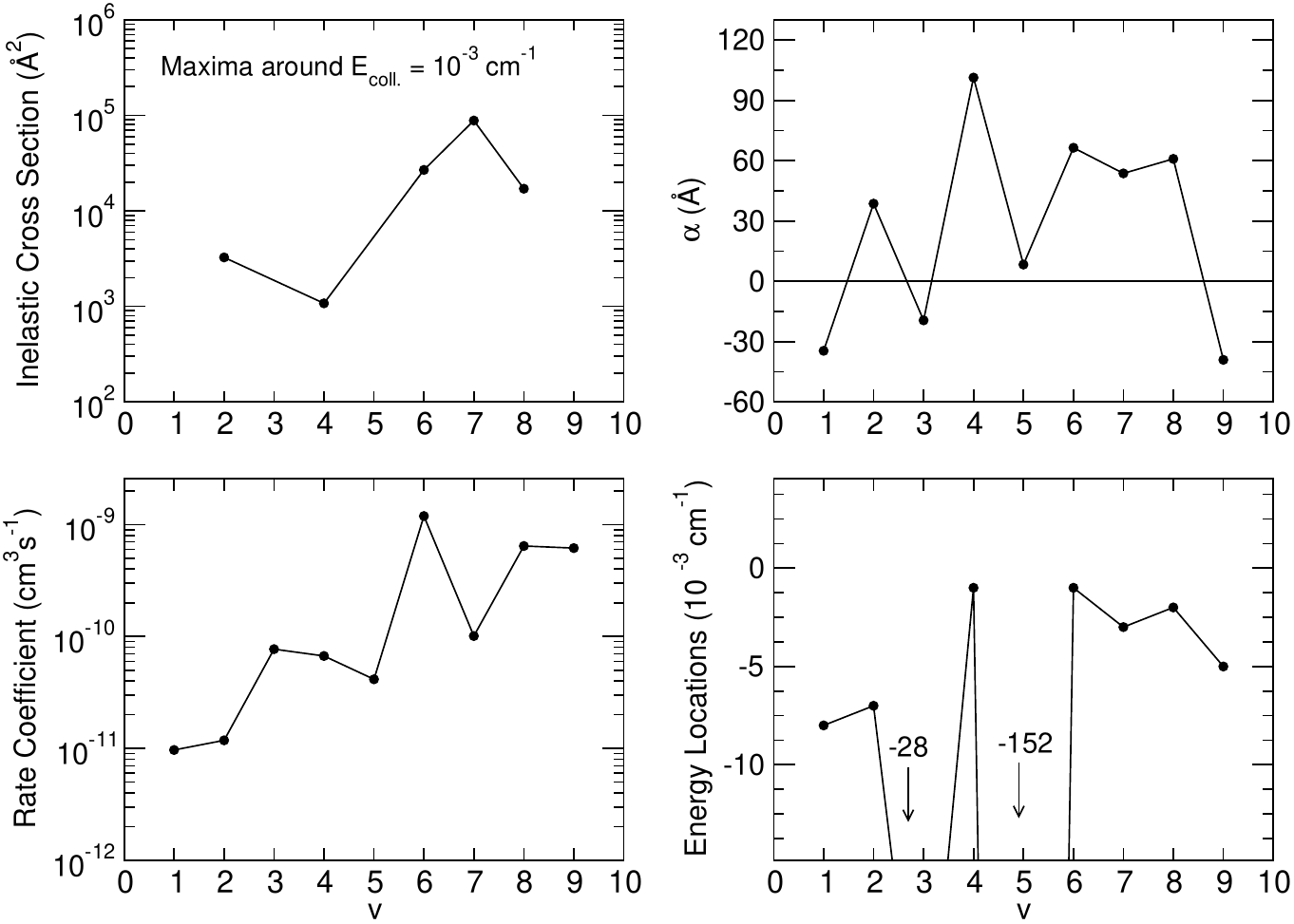}
\caption{Computed scattering attributes as a function of the initial
vibrational level preparation of the H$_2$ partner. Upper left
panel: total inelastic cross section maxima around the
10$^{-3}$ cm$^{-1}$  region. Upper right panel: sign and value of the
real part of the scattering length as a function of $v$. Lower left
panel: quenching rate coefficients from eq.(10). Lower right panel:
locations of bound and virtual states as s function of initial $v$
state.}\label{figure6}
\end{center}
\end{figure}

From the data in the figure one sees that the rates in the Wigner's
regime region show a constant increase as $v$ increases, thus
suggesting, as already found in our earlier work on neutral systems
\cite{11}, that when molecular targets have a low density of
vibrational states per unit of energy the corresponding quenching
efficiency has a strong dependence on the state preparation of the
molecule. We further see that the $v$=6 state shows a very strong
rate increase which is larger than that expected for it from the
general trend of the values in the panel. This initial state
exhibits both the presence of a shape resonance (see fig.4) and a
fairly large and positive scattering length, a fact which suggests a
bound state of the complex very near threshold, as confirmed by the
data in the lower right panel of the same figure. Thus, one may argue that ionic
interactions which are able to support metastable bound states for the triatomic complex
could also provide increased quenching efficiency: this feature, in
fact, is confirmed in figure 6 by the results provided by the H$_2$ state
preparations in the  $v$= 6, 7 and 8 levels.
We also note that negative scattering length values, i.e. signatures of
virtual state formation, are present for $v$=1, 3 and 9 H$_2$ states, exactly as indicated by
the RT mimimum features of figure 3. 

\section{Summary and Conclusions}
In the present analysis we have extended our previous work \cite{14} on the vibrational
collisional quenching processes at low and ultralow energies for the
subreactive channels of an ionic system, the Li$^++$H$_2(v)$ case, to
see if that collisional efficiency can be related both to features
of a realistic interaction potential from ab initio calculations
and to the specific internal energy content with which the target
molecule is being prepared before undergoing a collisional quenching.

One important element at the millikelvin regime has been found to be
the presence of non \emph{s}-wave contributions to the opacity functions of
the inelastic state-to-state vibrational channels. In other words,
substantial $l$=1 contributions appear when specific vibrational
initial states are considered, thereby causing shape resonances to
occur and a marked increase of the quenching efficiency to appear.
This is an important result which suggests that the reaction studied
could be controlled via selecting specific initial states in the
neutral molecule which in turn drive either up or down in size the
corresponding vibrational quenching rates.

Furthermore, our study has shown that the presence of virtual states
near threshold supported by the interaction potential is reflected
at the millikelvin regime by the appearance of very marked elastic
cross section minima which should be amenable to experimental
observation and should then allow us to test the quality of the
chosen potential energy surfaces, possibly modifying them according to
experiments.

Finally, the quenching rates at vanishing collision energies, i.e.
well into the Wigner's  regime of scattering cross sections,
indicate a strong dependence of their size on the selected initial
state of H$_2$: they turn out to vary by more than two orders of
magnitude  when moving from the $v$=1 initial state to the $v$=9
state. The main cause of changes is seen to 
be related to the increased target size as $v$ increases,
together with the additional appearance of either bound or virtual states
of the compound system at energy very close to threshold.
Both are important factors for driving upwards the quenching
efficiency, as illustrated by the numbers of Table 1 and by the
panels of figure  6.

In conclusion, the present calculations for an ionic case in which
both rotational and vibrational degrees of freedom of the molecular
partner are described as coupled to the incoming ion via a fairly
realistic PES \cite{13,14} indeed show a marked dependence of
scattering observables on the initial rovibrational state in which
the target molecule is being prepared and therefore suggest that the
latter state-preparation could be providing a viable procedure for
controlling the efficiency of the quenching process.

\begin{acknowledgments}
The financial support of  the University of Rome Research Committee,
of the CASPUR Computing Consortium, of the MUIR PRIN 2006 research
project is gratefully acknowledged.
\end{acknowledgments}

\listoffigures
\end{document}